\documentclass[aps,prl,twocolumn,groupedaddress,showpacs]{revtex4}
\usepackage{graphicx}
\draft

\begin{document}
\title{Quantum Hall Effect in Thin Films of Three-Dimensional Topological Insulators}
\author{Huichao Li$^1$}
\author{L. Sheng$^{1}$}
\email{shengli@nju.edu.cn}
\author{D. Y. Xing$^1$}
\email{dyxing@nju.edu.cn}

\address{$^1$ National Laboratory of Solid State Microstructures and
Department of Physics, Nanjing University, Nanjing 210093, China}

\begin{abstract}
We show that a thin film of a three-dimensional topological
insulator (3DTI) with an exchange field
is a realization of the famous Haldane model for quantum Hall
effect (QHE) without Landau levels. The exchange field
plays the role of staggered fluxes on the honeycomb lattice, and
the hybridization gap of the surface states is equivalent to
alternating on-site energies on the AB sublattices. 
A peculiar phase diagram for the QHE
is predicted in 3DTI thin films under an applied magnetic field, 
which is quite different from that either in
traditional QHE systems or in graphene.

\end{abstract}

\mbox{}\\

\maketitle

{\it Introduction.--}Haldane was the first to realize 
that quantum Hall effect (QHE) can
happen in systems without Landau levels (LLs). He proposed a
spinless electron model~\cite{Haldane} defined on a two-dimensional
(2D) honeycomb lattice with staggered magnetic fluxes and
alternating on-site energies on the AB sublattices. In the model,
the electrons retain their usual Bloch state character, but exhibit a
nonzero quantized Hall conductivity $\sigma_{xy}$ by simulation of
the ``parity anomaly'' of ($2+1$)D field theories, now termed as
quantum anomalous Hall effect. The Haldane model breaks
time-reversal symmetry. Kane and Mele~\cite{KaneMele} suggested that
the single-atomic-layer graphene with intrinsic spin-orbit coupling
constitutes two copies of the Haldane model. Since the two spin
components are mutually conjugate under time reversal, the Kane-Mele model
restores the time reversal symmetry. While the two spin components
contribute oppositely to the charge Hall conductivity, the
model displays a novel quantum spin Hall effect (QSHE).
The QSHE has been experimentally observed in quantum wells of mercury telluride~\cite{SHE0},
instead of graphene. A great deal of
research interest in the QSHE in recent years
has led to the birth of topological insulators,
which include a class of 2D and 3D band insulators with nontrivial
band topology~\cite{topoI}.

A 3D topological insulator~\cite{Kane0,TI0} (3DTI) is featured with the presence
of conducting surface states in the bulk band
gap, which are topologically protected and insensitive to impurity
scattering. On an isolated surface of the 3DTI, the energy
spectrum of the
surface states constitutes a single 
massless spin-momentum-locked Dirac cone~\cite{TI1},
which offers a unique platform to realize some exotic
relativistic quantum phenomena, such as
Majorana fermions~\cite{Major}.
In a perpendicular magnetic field $B_0$, the surface states
are quantized into LLs with energies
$E_{n} = \mathop{\rm sgn}(n) \sqrt {2e
B_0v_F^2\vert n\vert}$ (in $\hbar=c=1$ units), 
where $n=0,\pm 1,\pm 2\cdots$ and $v_{F}$ is the
Fermi velocity. Due to the existence of the
zero-mode ($n=0$) LL, which is particle-hole symmetric, 
the Hall conductivity is expected to
be half-integer quantized~\cite{Kane0}
$\sigma_{xy}=\nu\frac{e^2}{h}$, where $\nu=\ell+\frac{1}{2}$ with $\ell$
being an integer. 
Exciting experimental progress has been made recently. 
Cheng $et$ $al.$ and Hanaguri $et$ $al.$~\cite{LL0} detected  
the surface states in Bi$_2$Se$_3$ to form LLs in the presence of an applied 
magnetic field. Very recently, Br\"{u}ne $et$ $al.$~\cite{LL1} carried out a
Hall conductivity measurement in a high-quality 
strained HgTe layer, and observed a sequence of Dirac-like 
Hall plateaus,
which was ascribed to the topological surface states. 
It was also suggested theoretically
that electrons in the surface states of a 3DTI with 
exchange splitting could give rise to a QHE
in the absence of an applied magnetic field, i.e.,
the quantum anomalous Hall effect~\cite{qahe}.
For a thin film of a 3DTI, the finite-size confinement will mix the
surface states at the top and bottom surfaces and create a hybridization energy
gap in the spectrum of the surface states~\cite{hybgap0,hybgap1}. The hybridization
gap can be controlled by tuning the thickness of the 3DTI
film~\cite{hybgap0,hybgap1}. The exchange field (or equivalently
a Zeeman energy) and the hybridization gap effectively
add different types of mass to the Dirac
fermions~\cite{mass}
and provide ways to control the properties of the surface states, but
their effect on the
QHE has not been well understood.

The Haldane's QHE model~\cite{Haldane} has been difficult to be
realized in condensed matter systems, though Shao $et$
$al.$~\cite{ColdAtom} suggested recently that it could be simulated
by using ultracold atoms.  In this
Letter, we propose that a thin film of a 3DTI in the presence of an
exchange field is a natural realization of the Haldane model. 
The exchange field plays the role of the staggered fluxes in the Haldane
model, and the hybridization gap is
equivalent to the alternating on-site energies on AB sublattices.
Under an applied magnetic field, a 3DTI thin film and 
the Haldane model are still equivalent in Hamiltonian form, so that we can
study the QHE in the 3DTI film conveniently based upon the Haldane model. 
The other purpose of this Letter is to report a peculiar phase
diagram for the QHE in thin films of 3DTIs, due to the interplay
between Zeeman energy $g$ and hybridization gap $\Delta$. 
For $\Delta=0$, a nonzero $g$ only shifts the positions of the LLs
without creating new Hall plateaus. If $g=0$ but $\Delta\neq 0$, a
new $\nu=0$ plateau will appear, in addition to the original
$\nu=2(\ell+\frac{1}{2})$ odd-integer plateaus. The simultaneous
presence of nonzero $g$ and $\Delta$ causes splitting of the
odd-integer Hall plateaus, and so all integer ($\nu=\ell$) plateaus
emerge. More interestingly, as the product of $g$ and $\Delta$ is
equal to certain critical values, the split plateaus can merge
again, and most odd-integer (or even-integer) plateaus disappear.

{\it Thin Films of 3DTIs.--}We start with an effective Hamiltonian
for electrons in the surface states of a 3DTI thin 
film~\cite{hybgap1} with an exchange energy~\cite{qahe}
\begin{equation}
 H\left( k \right) = -Dk^2+{v_F}\left( {{k_y}{{\hat \sigma }_x} - {k_x}{{\hat \sigma }_y}}
 \right) + (\frac{\Delta }{2} - B{k^2}){\tau _z}{\hat \sigma _z} + g{\hat \sigma
 _z}.
\label{eq:ham0}
\end{equation}
Here, the first two terms are the Hamiltonian for two isolate
surfaces with ${v_F}$ the Fermi velocity, ${\bf k}$ is the momentum with
respect to Dirac point $\Gamma$, and $\hat{\sigma}_\alpha$ are the
Pauli matrices for electron spin with $\alpha=x,y,z$. The third term
stands for the coupling between the upper and lower surface states
of the thin film with ${\tau}_z =1$ $(-1)$ representing the bonding
(antibonding) between them, and the last one is the exchange field
with strength $g$.
We will consider the case, where the condition $B^2-D^2 > 0$ for 
band inversion is satisfied~\cite{hybgap1}.
Since Hamiltonian (1) is block diagonal for $\tau_z=\pm 1$, the
Chern numbers $C_{\tau_z}$ of the occupied
valence bands with $\tau_z=+1$ and $-1$ can be
calculated separately, yielding $ C_{{\tau _z}} =-\tau_z\left[{\mathop{\rm
sgn}} (B) + {\mathop{\rm sgn}} (\Delta  + 2g{\tau _z})
\right]/{2}$. The total Chern number $C=C_{+}+C_{-}$ is
given by
\begin{equation}
C=\frac{1}{2} [{\mathop{\rm
sgn}} (\Delta-2g) - {\mathop{\rm sgn}}(\Delta + 2g)]\ .
\end{equation}
We note that even though the ${\tau _z}$-dependent Chern numbers
depend on the sign of parameter $B$, the total Chern number is
independent of $B$.  In Fig.\ 1, we show the phase diagram of the
3DTI thin film in the presence of exchange splitting. Lines
$g=-\Delta /2$  and $g=-\Delta /2$ are the critical boundaries
between the normal insulator with $C=0$ and the Chern insulator with
$C=\pm 1$, the latter being in the lower and upper shaded regions.
In the regions of $C=\pm 1$, quantum anomalous Hall effect with
$\sigma_{xy}=\pm\frac{e^2}{h}$ occurs.

\begin{figure}[h]
    \centering
    \includegraphics[clip,width=2.2in, angle=0]{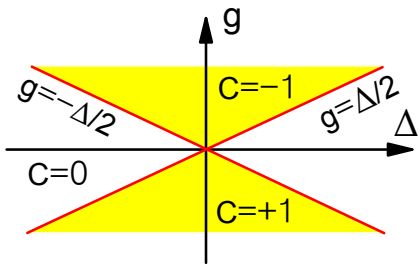}
    \caption{Schematic phase diagram on the hybridization potential $\Delta $
     and exchange energy $g$ plane, in which two solid lines $g=\pm\Delta/2$
      separate four phase regions with $C=\pm 1$ and $C=0$. }
    \label{fig:fig0}
\end{figure}

{\it Haldane Model.--}The Hamiltonian of the Haldane model reads~\cite{Haldane}
\begin{equation}
H =- \sum\limits_{\langle i,j\rangle} {tc_i^\dag {c_j}}
+ \sum\limits_{\langle\langle i,j\rangle\rangle} {{t_2}{e^{i\phi_{ij}}}c_i^\dag {c_j}}
+ \sum\limits_i {{V_i}c_i^\dag {c_i}}\ .
\end{equation}
Here, $c_i^\dag $ (${c_i}$) is the fermion creation (annihilation)
operator at site $i$, $t$ ($t_2$) is the hopping integral between
the nearest-neighbor (next-nearest-neighbor) sites $i$ and $j$, and
${V_i} = \pm M$ for sublattices $A$ and $B$, respectively.
$\phi_{ij}=\pm \phi$ is the accumulated Peierls phase from site $i$
to its second neighbor $j$, due to a staggered magnetic-flux
density, where the positive sign is taken for an electron hopping along
the arrows indicated in the Fig.\ 1 of Ref.\ [1].  It has been
shown~\cite{Haldane} that the Haldane model (3) exhibits a quantized
Hall conductivity $\pm \frac{e^2}{h}$ for $3\sqrt{3}\vert
t_2\sin\phi\vert>\vert M\vert$, and behaves as a normal insulator otherwise.

The low-energy behavior of electrons in the Haldane model is
governed by the dispersion relation near the two inequivalent Dirac
points $K$ and $K'$. We can expand the Hamiltonian to the linear
order in the relative momentum ${\bf k}$ with respect to the two
Dirac points, and obtain
\begin{eqnarray}
H\left( k \right)=v_F
\left( {{\tau_z}{k_y}{\hat{\sigma}_x} - {k_x}{\hat{\sigma}_y}} \right)
+\frac{\Delta}{2}\hat{\sigma}_z+g{\tau}_z{\hat{\sigma}
_z}\ .
\end{eqnarray}
Here, ${v_F} = {{3at}}/{{2}}$, ${\Delta }/{2} = M$, $g = - 3\sqrt 3
{t_2}\sin \phi$, and a constant energy $-3{t_2}\cos \phi$ has been
omitted. $\hat{\sigma}_{\alpha}$ represent two sublattices and
${\tau}_{z}=\pm 1$ correspond to two Dirac valleys. Using a unitary
transformation $H'={U^\dagger}HU $ in Eq.\ (4) with $U=\left[ {1 +
{\hat{\sigma}_y} + \left( {1 - {\hat{\sigma}_y}} \right){\tau _z}}
\right]/2$, we obtain $H'\left( k \right) = {v_F}\left( {{k_y}{{\hat
\sigma }_x} - {k_x}{{\hat \sigma }_y}} \right) + \frac{\Delta
}{2}{\tau _z}{\hat \sigma _z} + g{\hat \sigma _z}$. To the linear
order of $k$, $H'(k)$ is identical in form to Eq.\ (1), even though
parameters in the two Hamiltonians have different physical meanings.
More importantly, the condition for the occurrence of the quantum
anomalous Hall effect in the Haldane model~\cite{Haldane},
$3\sqrt{3}\vert t_2\sin\phi\vert>\vert M\vert$, is also identical to
that for a 3DTI thin film, i.e., $\vert g\vert>\vert\Delta\vert/2$
as shown in Fig.\ 1,
due to the mapping of parameters below Eq.\ (4). Therefore, a 3DTI
thin film in the presence of an exchange energy is a natural
realization of the Haldane model. The exchange energy plays the role
of the staggered fluxes in the Haldane model, and the hybridization
gap of the surface states is equivalent to the alternating on-site
energies.

{\it Quantum Hall Effect of 3DTI Thin Films.--}We now study
the QHE in a thin film of a nonmagnetic 3DTI.
As a perpendicular magnetic field $B_0$ is applied to the film,
the action on the spin degrees of freedom is the Zeeman energy, which can
be described by the last term, $g\sigma_z$ in Eq.\ (1) (with
$g=g_{eff}\mu_{B}B_{0}$); and the action on
the orbital motion can be included by use of the Peierls substitution
${\bf k}\rightarrow ({\bf k}-e{\bf A})$ with  ${\bf A}$
the vector potential due to $B_0$.
Since the unitary transformation $U$ introduced above does not affect the 
orbital degrees of freedom, it is apparent that 
Hamiltonian (1) is still equivalent to the
Haldane model given by Eqs.\ (3) and (4) in this case,
provided that the same vector potential
${\bf A}$ is included into these equations. As a result, the
QHE in a thin film of a 3DTI is equivalent to
that in the Haldane model. The
LLs in the Haldane model Eq.\ (4) subject to a magnetic field
have been solved in the original work of
Haldane~\cite{Haldane}
\begin{equation}
{E_{{\tau _z},n}} =  {\mathop{\rm sgn}}(n) \sqrt {
w_{1}^2\vert n\vert + {{\left( {\frac{\Delta }{2} + g{\tau _z}}
\right)}^2}},
\end{equation}
for nonzero integer $n$, and
\begin{equation}{E_{{\tau _z},0}} = [g+ (\Delta/2) \tau _z]\mbox{sgn}(eB_0)\ ,
\end{equation}
for $n=0$. Here,  $w_1=\sqrt{2\vert e
{B_0}}\vert v_F$ is the width of the $\nu=1$ Hall plateau
at $g=\Delta=0$. When both $g$ and $\Delta$ vanish, Eqs.\ (5) and (6)
reduce to the standard LLs for massless Dirac fermions,
which are additionally degenerate for $\tau_z=\pm 1$, i.e., ${E_{ + ,n}} = {E_{ -
,n}}$. Nonzero $g$ and $\Delta$
may shift the relative positions of the LLs and cause the LLs
to split, which will determine the
quantization rule of the Hall conductivity.

The Hall conductivity will be calculated numerically from
Eq.\ (3)  by setting
$\phi={\pi}/{2}$ in a system of size $64 \times 64$. The vector potential
of the applied magnetic
field is introduced into Eq.\ (3) via the integral form of the
Peierls substitution $t\rightarrow te^{i\theta_{ij}}$ ($t_2\rightarrow
t_2e^{i\theta_{ij}}$), where ${\theta
_{ij}} = e\int_i^j
{\bf A} \cdot d{\mbox{\boldmath{$l$}}}$ with the integral being
along the electron hopping path. The magnetic flux
per hexagon is given by $\varphi=\sum_{ij} {{\theta _{ij}}}  = \pi
{B_0}3\sqrt 3 {a^2}/{\phi _0} = 2\pi /M_0$, where the summation runs
over six links around a hexagon, ${\phi_0} =h/e$ is the flux
quantum, and $M_0$ is an integer.
The Hall conductivity at zero temperature can be calculated by use
of the standard Kubo formula
\begin{equation}{\sigma _{xy}} = \frac{{4{e^2}}}{{S}}
\sum\limits_{{\varepsilon _{m}}<{E_F}<{\varepsilon _{n}}}
 {\frac{\mbox{Im}({\langle
n|{v_x}\left| m \right\rangle \langle m|{v_y}\left| n \right\rangle})}
{{{{\left( {{\varepsilon _m} - {\varepsilon _n}}
\right)}^2}}}}\ .
\end{equation}
Here, ${\varepsilon _{m}}$ and ${\varepsilon _{n}}$ are the
eigenenergies corresponding to occupied state $\left| {m}
\right\rangle$ and empty state $\left| {n} \right\rangle$,
respectively.
$S$ is the area of the hexagon lattice,  and ${v_\alpha }$ is the
velocity operator with $\alpha = x$ or $y$.
With tuning chemical potential
$E_F$, a series of integer-quantized plateaus of ${\sigma
_{xy}}$ will appear, each one
corresponding to $E_F$ moving in the gaps between two neighboring LLs.
When $g =\Delta= 0$, Eq.\ (3) reduces to the tight-binding
model for graphene without spin degrees of freedom. It follows that the
Hall conductivity in the 3DTI thin film is half of
that in graphene. The Hall
conductivity for the 3DTI thin film is thus odd-integer quantized
${\sigma _{xy}} = 2(\ell +\frac{1}{2})\frac{e^2}{h}$ with the prefactor 2
originating from two surfaces, as expected.

\begin{figure}[h]
    \centering
    \includegraphics[clip,width=3.0in, angle=0]{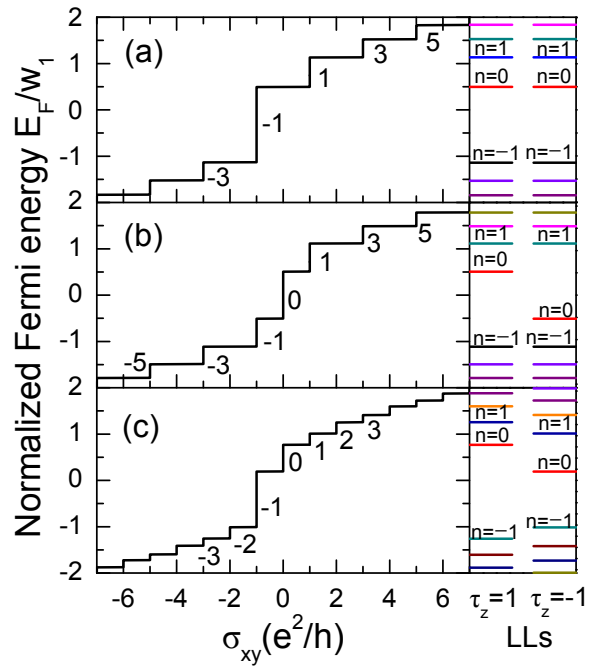}
    \caption{Calculated Hall conductivity ${\sigma _{xy}}$ in
    units of ${e^2}/{h}$ as a function of ${E_F}$
    with magnetic flux $\varphi  = 2\pi /1024$
    for (a) $g = 0.5w_{1}$ and $\Delta  =
    0$, (b) $g = 0$ and $\Delta  = w_{1}$,
     and (c) $g = 0.5w_{1}$ and $\Delta  = 0.6w_{1}$.  The corresponding LL energies
     for ${\tau _z} =  \pm 1$ are shown in the right panel.}
    \label{fig:fig1}
\end{figure}
\begin{figure}[h]
    \centering
    \includegraphics[clip,width=3.0in, angle=0]{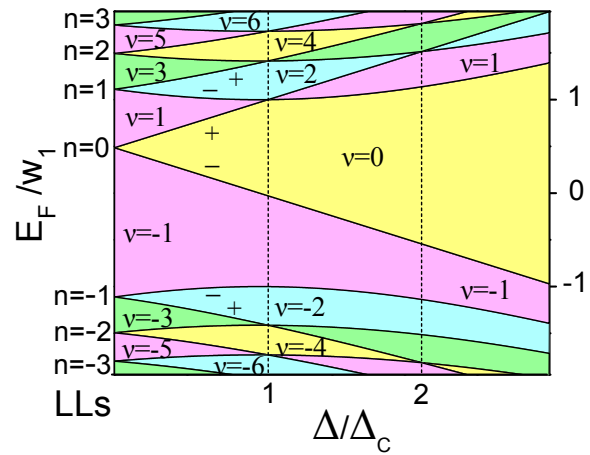}
    \caption{Phase diagram for the QHE on the $E_F/w_1$
    versus $\Delta/\Delta_c$ plane.
    The phase boundary is determined by the energies of the LLs
    as functions of $\Delta/\Delta_c$, calculated from
    Eqs.\ (5) and (6). Here, $\Delta_c=w_{1}^2/2g$, and the
    Zeeman energy is taken to be $g=0.5w_1$.}
    \label{fig:fig2}
\end{figure}

In Fig.\ 2, the calculated Hall conductivity ${\sigma _{xy}}$ in units of
$\frac{e^2}{h}$ is plotted as a function of $E_F/w_1$ with magnetic
flux $\varphi = 2\pi /1024$ for some different values of $g$ and
$\Delta$. From Eqs.\ (5) and (6), we can see that for $g \ne 0$ and
$\Delta = 0$, while their positions shift with $g$, all the
LLs are still degenerate for $\tau_z=\pm 1$,
so that the total Hall
conductivity remains to be odd-integer quantized ${\sigma _{xy}} =
(2\ell + 1)\frac{e^2}{h}$, as shown in Fig.\ 2(a). 
For $g = 0$ and $\Delta \ne 0$, the LLs with
$n\ne 0$ are degenerate for ${\tau _z}=\pm 1$, but there is a
splitting of the $n=0$ LL, yielding a new plateau of ${\sigma _{xy}}
= 0$, as shown in Fig.\ 2(b). For $g \ne 0$ and $\Delta \ne 0$, the
additional degeneracy of all the LLs is generally lifted, and each
degenerate LL splits apart into two so that all integer Hall
plateaus ${\sigma _{xy}} = \ell\frac{e^2}{h}$ appear, as shown in
Fig.\ 2(c).

In Fig.\ 3, we plot the energies of the LLs calculated from Eqs.\
(5) and (6) as functions of $\Delta$ for a fixed Zeeman energy,
which effectively determine a phase diagram for the QHE on the
$E_{F}$ versus $\Delta$ plane. The splitting of each LL increases
with $\Delta$, which can be understood from Eqs.\ (5) and (6). Interestingly,
when $\Delta$ is integer multiples of certain critical value
$\Delta_c$, the LLs for $\tau_z=\pm 1$ and different $n$ cross each
other at the same time, leading to disappearance of nearly half of
the Hall plateaus. For example, $E_{+, n}= E_{-, n+1}$ at $\Delta
=\Delta_c$, $E_{+, n}= E_{-, n+2}$ at $\Delta =2\Delta_c$, and on
analogy of this, for $n\ge 0$. The LLs for $n<0$ cross in a similar
manner. It is straightforward from Eq.\ (5) to obtain $g\Delta_c  =
w_{1}^2/2$ or $\Delta_c=w_{1}^2/2g$. The Hall conductivity as a
function of $E_F$ is shown in Fig.\ 4(a) at $\Delta =\Delta_c$ and
4(b) at $\Delta =2\Delta_c$, respectively. In the former, the Hall
conductivity $\sigma_{xy}=\nu\frac{e^2}{h}$ is quantized into
even-integer $\nu=2\ell$ plateaus plus a single odd-integer $\nu=-1$
plateau, and in the latter, it displays odd-integer $\nu=(2\ell+1)$
plateaus plus two even-integer $0$ and $-2$ plateaus. In both cases,
the split Hall plateaus merge again partly, and most odd-integer
[even-integer] plateaus disappear in Fig.\ 4(a) [4(b)].

\begin{figure}[h]
    \centering
    \includegraphics[clip,width=3.0in, angle=0]{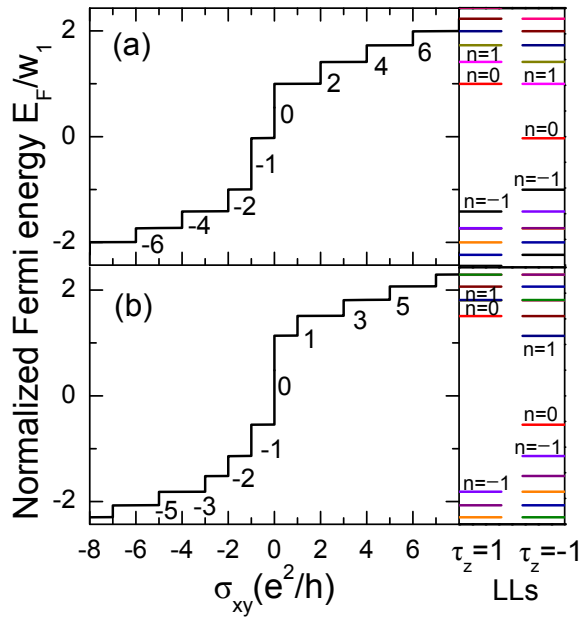}
    \caption{Calculated Hall conductivity ${\sigma _{xy}}$  in units of ${e^2}/{h}$
    as a function of ${E_F}$ with magnetic flux
    $\varphi  = 2\pi /1024$ for (a) $g = 0.5w_{1}$ and $\Delta  = w_{1}$,
    and (b) $g = 0.5w_{1}$ and $\Delta  = 2w_{1}$. The corresponding LL energies
     for ${\tau _z} =  \pm 1$ are shown in the right panel. }
    \label{fig:fig3}
\end{figure}

{\it Summary.--} In summary, we have shown that the Haldane model
for the QHE without LLs can be realized in condensed matter systems
by use of a 3DTI thin film with an exchange field. As a
perpendicular magnetic field is applied to the 3DTI thin film,
Hamiltonian (1) and Eq.\ (4) for the Haldane model are still
equivalent to each other, provided that the vector potential {\bf A}
is included in both equations. A rich phase diagram for the QHE in
the 3DTI thin film is predicted as a consequence of the interplay
between $g$ and $\Delta$. The simultaneous presence of nonzero $g$
and $\Delta$ causes splitting of original odd-integer Hall plateaus
$\sigma_{xy}=(2\ell+1)\frac{e^2}{h}$ into
$\sigma_{xy}=\ell\frac{e^2}{h}$. Remarkably, when the product of $g$
and $\Delta$ is at certain critical values, the plateaus can merge
again partly, and most odd-integer (or even-integer) plateaus
disappear.

{\it Acknowledgments.--}This work is supported by the State Key
Program for Basic Researches of China under Grant Nos. 2009CB929504,
2007CB925104 (LS), 2011CB922103 and 2010CB923400 (DYX), the
National Natural Science Foundation of China under Grant Nos.
10874066 and 11074110, and the Science Foundation of Jiangsu Province in China
under Grant No. SBK200920627.

%

\end{document}